\documentstyle[12pt,dina4]{article}

\newcommand{\bc}{\begin{center}}
\newcommand{\ec}{\end{center}}
\newcommand{\bt}{\begin{tabular}}
\newcommand{\et}{\end{tabular}}

\newcommand{\be}{\begin{equation}}
  \newcommand{\ee}{\end{equation}}
\date{}
\begin{document}
\title{Medium effects in the production and decay of $\omega$- and
$\rho$-resonances in pion-nucleus interactions\footnote{Supported by DFG}}

\author{ W. Cassing$^1$, Ye.S. Golubeva$^2$, A.S. Iljinov$^2$  \\
and L.A. Kondratyuk$^3$}
\maketitle
\bc$^1$ Institute  for Theoretical Physics, University of
Giessen, D-35392 Giessen, Germany \\

$^2$ Institute of Nuclear Research, 117312 Moscow, Russia
\\

$^3$ Institute of
Theoretical and Experimental Physics, 117259
Moscow, Russia
\ec

\begin{abstract}
The $\omega$- and $\rho$-resonance production and their dileptonic 
decay in  $\pi^- A$ reactions at GSI energies are calculated within 
the intranuclear cascade (INC) approach.  The invariant mass 
distribution of the dilepton pair for each resonance is found to have 
two components which correspond to the decay of the resonances 
outside and inside the target nucleus. The latter components are 
strongly distorted by the nuclear medium due to resonance-nucleon 
scattering and a possible mass shift at finite baryon density. These 
medium modifications are compared to background sources in the 
dilepton spectrum from $\pi N$ bremsstrahlung and the Dalitz decays 
of $\Delta$'s, $\omega$ and $\eta$ mesons produced in the reaction.
\end{abstract}

\newpage
The question about the properties of hadronic resonances in the 
nuclear medium has received a vivid attention during the last years 
(cf. Refs. \cite{1,2,3,4}). Here, QCD inspired effective Lagrangian 
models \cite{1,2} or approaches based on QCD sum rules \cite{3,4} 
predict that the masses of the vector mesons $\rho$, $\omega$ and 
$\phi$ should decrease with the nuclear density. Furthermore, along 
with a dropping mass the phase space for the resonance decay also 
decreases which results in a modification of the resonance width in 
matter.  On the other hand, due to collisional broadening - which 
depends on the nuclear density and the resonance-nucleon interaction 
cross section ( cf. Refs. \cite{5,6}) - the resonance width should 
increase again.

The in-medium properties of vector mesons have been addressed 
experimentally so far by dilepton measurements at the SPS, both for 
proton-nucleus and nucleus-nucleus collisions 
\cite{CERES,HELIOS,HELI}. As proposed by Li {\it et al.} \cite{Li}, 
the enhancement in $S + Au$ reactions compared to $p + Au$ collisions 
in the invariant mass range $0.3 \leq M \leq 0.7$ GeV might be due to 
a shift of the $\rho$ meson mass. The microscopic transport studies 
in Refs. \cite{11,12} for these systems point in the same direction, 
however, also more conventional selfenergy effects cannot be ruled 
out at the present stage \cite{11,Wamb}. It is therefore necessary to 
have independent information on the vector meson proporties from 
reactions, where the dynamical picture is more transparent, i.e. in 
pion-nucleus collisions. Here, especially the $\omega$ meson can be 
produced with low momenta in the laboratory system, such that a 
substantial fraction of them will still decay inside a heavy nucleus 
\cite{Metag}.

The mass distributions of the vector mesons in the latter case are 
expected  to have a two component structure \cite{6} in the dilepton 
invariant  mass spectrum: the first component corresponds to 
resonances decaying in the vacuum, thus showing the free spectral 
function which is very narrow in case of the $\omega$ meson; the 
second (broad) component then corresponds to the resonance decay 
inside the nucleus. We will use that (in first order) the in-medium 
resonance can also be described by a Breit- Wigner formula with a 
mass and width distorted by the nuclear environment.

In this letter we present first microscopic calculations for  the 
production and dileptonic decay of $\omega$ and $\rho$ resonances in 
$\pi^- A$ collisions at pion momenta of 1.3 GeV/c available at GSI in 
the near future. The calculations are performed  within the framework 
of the intranuclear cascade model (INC) \cite{7} which was extended 
earlier \cite{8,9} to account for the in-medium resonance decays. 
First calculations for the shapes of the $\omega$ and $\rho$ peaks in 
proton (antiproton) induced reactions $p A\to VX\to e^+e^-X$ ($\bar p 
A\to VX\to e^+e^-X$) have been reported in Refs. \cite{8,9}.  Here we 
consider explicitly $\pi^-$ induced reactions and also compute the 
background sources in the dilepton spectrum from  nucleon-nucleon and 
pion-nucleon bremsstrahlung as well as the Dalitz decays $\Delta \to 
N e^+e^-$, $\omega \to \pi^0e^+e^-$ and $\eta \to \gamma e^+e^-$ 
following Refs. \cite{11,12,10}.

The vector mesons $\rho, \omega$ are produced in the first hard 
pion-nucleon collision in the target and propagate through the rest 
of the nucleus.  If the wave length of the produced meson is much 
smaller than the nuclear radius, it is possible to use the eikonal 
approximation for the Green function describing its propagation from 
the point $\vec{r}=(\vec{b},z)$ to $\vec{r}'=(\vec{b}',z')$ \cite{6},

\be
\label{Green}
G_k(\vec{b}',z';\vec{b},z)=\frac{1}{2ik} exp
\{i\int^{z'}_z[k+{\frac{1}{2k}(\Delta +4\pi f
(0)\rho(\vec{b}},\zeta)]d\zeta\}
\ee
$$
\times \delta(\vec{b}-\vec{b}')\theta(z'-z),
$$
where the $z$-axis is directed along the resonance momentum $\vec{k}$,
$\vec{b}$ is the impact parameter, $f(0)$ is the resonance--nucleon forward
scattering amplitude, $\rho$  is the nuclear density and $\Delta$ is the
inverse resonance propagator
\be
\Delta=P^2-M^2_R+iM_R\Gamma_R
\ee
with $M_R,\Gamma_R$ and $P$ being the mass, width and four momentum of the
resonance. The latter can be defined through the four momenta of its decay
products
\be
P=p_1+p_2+...
\ee

Let us now consider the production of the vector mesons $\rho$ and $\omega$
in the nuclear target and their decay into the $e^+e^-$ pair:
  \be \pi^- A \to RX\to e^+e^-X. \ee
If the resonance $R$ is created at  $(\vec{b},z)$ and decays in
$(\vec{b'},z')$, the exponent of the Green function (1) can be
separated into the contributions from the two regions

1) $z\leq z'\leq z_s= \sqrt{R^2-\vec{b}^2}$ and

2) $z'\geq z_s=\sqrt{R^2-\vec{b}^2}.$
\\
When the resonance decays inside the nucleus of radius $R$, only the first 
part contributes and the inverse resonance propagator has the form
\be
\Delta^*=\Delta+4\pi  f(0)\rho_0= P^2-M^{*2}_R+iM_R^*\Gamma_R^*
\ee
where
\be
M_R^{*2}=M^2_R-4\pi Re{f(0)}\rho_0,
\ee
\be
M_R^*\Gamma^*_R=M_R\Gamma_R+4\pi Im{f(0)}\rho_0.
\ee
If the resonance decays outside the nucleus, both regions contribute
and the probability for lepton pair production with total
momentum $\vec{P}$ and invariant mass squared $P^2$ can be
written as
\be
\label{both}
|M(\vec{P}, P^2)|^2=N|f_{\pi^-N \to RX}|^2
\int d^2\vec{b} dz \ \rho(\vec{b},z)|A_{in}(\vec{P},P^2; \vec{b},z)+
A_{out}(\vec{P}, P^2; \vec{b},z)|^2 .
\ee
In Eq. (\ref{both}) the contributions from the first and second part can be
written as
\be
A_{in}(\vec{P}, P^2; \vec{b},
z)=\frac{1-exp[i(\Delta^*/2k)(z_s-z)]}{\Delta^*},
\ee
and
\be
A_{out}=\frac{exp[i(\Delta^*/2k)(z_s-z)]}{\Delta},
\ee
where $f_{\pi^-N \to RX}$ is the amplitude for the resonance
production on a nucleon in the channel RX and $N$ is the normalization
factor which contains the branching ratio  $R\to e^+e^-$.

When the resonance decays inside the nucleus, its form is
described by the Breit--Wigner formula (9) with $\Delta^*$ from (5),
that contains the effects of collisional broadening
\be
\Gamma^*_R=\Gamma_R+\delta \Gamma ,
\ee
where
\be
\delta\Gamma = \gamma v\sigma_{(RN)}\rho_B,
\ee
and a shift of the meson mass
\be
M^{*}_R=M_R+\delta M_R ,
\ee
where
\be
\delta M_R=-\gamma v \sigma_{(RN)}\rho_B\alpha.
\ee
In Eqs. (12) and (14) $v$ is the resonance velocity with respect to 
the target at rest, $\gamma$ is the associated Lorentz factor, 
$\rho_B$ is the nuclear density, $\sigma_{(RN)}$ is the 
resonance--nucleon total cross section and $\alpha = (Re f(0))/(Im f(0)).$

If the ratio $\alpha$ is small - which is actually the case for the 
reactions considered because many reaction channels are open - the 
broadening of the resonance will be the main effect. The sign of the 
mass shift depends on the sign of the real part of the forward RN 
scattering amplitude which, in principle, also depends on the 
momentum of the resonance. For example, at low energy various authors  
\cite{1,2,3,4}  predict a decreasing mass of the vector mesons 
$\rho$, $\omega$ and $\phi$ with the nucleon density, whereas Eletsky 
and Ioffe have argued  recently \cite{13} that the $\rho$ might 
become heavier in nuclear matter at energies of  2-7 GeV.

In the following we consider the reaction $\pi^- A\to VX\to e^+e^-X$ 
for Pb-targets at a pion  momentum of 1.3 GeV/c.  The yields of 
$\omega$ -- and $\rho$-- mesons are calculated within the framework 
of the intranuclear cascade model developed  in Ref. \cite{7}, where 
the nucleus is devided into series of concentric zones and it is 
possible to follow the propagation of each produced particle from one 
zone to another. The decay of the resonances to $e^+e^-$ with their 
actual spectral shape was performed here as in  Refs. \cite{8,9}.  
The calculation proceeds as follows:  when the resonance decays into 
dileptons inside the nucleus its mass is generated  according to the 
Breit--Wigner distribution with $M^*_R=M_R+\delta M_R$ and 
$\Gamma^*_R=\Gamma_R+\delta \Gamma_R$, where the collisional 
broadening  and the mass shift are calculated according to the local 
nuclear density. Its decay to dileptons is recorded as a function of 
the corresponding invariant mass bin and the local density.  If the 
resonance leaves the nucleus, its spectral function automatically 
coincides with the free distribution because $\delta M_R$ and $\delta 
\Gamma_R$ are zero in this case (cf. Eqs. (12), (14)).

For the $\omega$ production we consider the channels:  $\pi^- p\to 
\omega n$ and $\pi^-n \to \omega \pi^-n$ with parametrizations from 
Ref. \cite{14}.  Though the momentum of 1.3 GeV/c is smaller than the 
threshold momentum for  the second channel, it contributes to  
$\omega$ production in  $\pi^-A $ collisions due to the Fermi motion 
of the target nucleons. For $\rho$-meson production we use the same 
cross sections as for $\omega$-mesons; this holds experimentally 
within 20\%.

The resulting momentum distribution of all decaying $\omega$-mesons 
and those decaying inside  the nucleus (at densities $ \rho \geq 
0.03\rho_0$) are displayed in Fig. 1 by the solid and dotted 
histograms, respectively. They  extend to $1.1$ $GeV/c$ with average 
momenta of about $0.65$ and $0.53$ $GeV/c$, respectively. The bump in 
the full spectrum around 0.5 GeV/c describes the contribution of the 
channel $\pi^-n \to \omega \pi^-n$. Using a cut of $0.03\rho_0$, 
approximately 47\% of the produced $\omega$- mesons are absorbed at 
finite density. We note that the $\omega$-mesons decaying  outside 
the nucleus show a momentum distribution very similar to the solid 
curve in Fig. 1 especially at high laboratory momenta. On the other 
hand, $\omega$ mesons with low momenta to a large extend still decay 
within the target nucleus.

In Fig. 2a  we present the $e^+e^-$ mass spectrum from $\omega$-
decays neglecting collisional broadening, however, including a
medium dependent $\omega$-mass as suggested by Hatsuda and Lee
\cite{3}
\be
\label{Brown}
M^*_R=M_R(1 - 0.18 \rho_B(r) / \rho_0),
\ee
where $\rho_B (r)$ is the nuclear density at the resonance decay and 
$\rho_0 = 0.16 fm^{-3}$. Fig. 2a describes the full mass distribution 
whereas Fig. 2b includes a cut in the $\omega$ momentum for $p_{lab} 
\leq 0.4 GeV/c$.  The shaded histograms describe the contributions of 
the '$in$' components, while the solid histograms are the sum of the 
'$in$' and '$out$' components.  Due to sizeable surface effects (with 
lower nucleon density) the mass distribution of the '$in$' components 
is spread out. Nevertheless, we still observe two sharp peaks 
corresponding to the in-medium and vacuum masses, respectively.  As 
expected from Fig. 1, a cut in the momentum distribution for $p_{lab} 
\leq$ 0.4 GeV/c enhances the contribution from the '$in$' component 
relative to the '$out$' component (Fig. 2b).

In Fig. 2c  we show the dilepton spectrum from $\omega$
decays when  a mass shift and collisional broadening
are taken into account. Here we have parametrized the
momentum dependent $\omega-N$ cross section by
\be
\sigma_{\omega N} (p_{lab}) = A + B/ p_{lab}
\ee
with  $A=11 mb$ and $B=9 mb*GeV/c$, which is taken as a guide due to 
the lack of experimental data\footnote{ The respective cross section 
in ref. \cite{15} is too large by about a factor of 2.}. Furthermore, 
we have assumed  $\sigma_{\rho N} \approx \sigma_{\pi N}$ in our 
calculations for $\rho + N$ collisions.  Again the relative 
contribution of the '$in$' component (shaded histograms) in the 
$e^+e^-$ invariant mass spectrum considerably increases when 
$\omega$-mesons with momenta less than 0.4 GeV/c are selected (Fig. 
2d). Due to collisional broadening the width of the '$in$' component 
increases substantially relative to the free $\omega$ width such that 
a pronounced peak for the '$in$' component can no longer be observed.

Whereas the $\omega$ dilepton mass distributions show quite 
considerable changes for the different scenarios proposed, it is 
questionable if these modifications actually can be seen when 
including all other known sources of dilepton channels, i.e. 
$\pi$-nucleon and proton-neutron bremsstrahlung, the Dalitz decays of 
the $\eta, \omega$ and $\Delta$ as well as the direct decay of the 
$\rho^0$ meson. The individual branching ratios and formfactors for 
the latter processes are taken from Ref. \cite{11,12}, while the 
bremsstrahlung channels are evaluated in the soft photon 
approximation \cite{10}, which will provide an upper estimate for the 
dilepton yield from bremsstrahlung.

In Fig. 3 we show the inclusive dilepton spectrum for $\pi^-$ + Pb at 
1.3 GeV/c - in logarithmic representation - from the various channels 
described above including collisional broadening, however, no mass 
shifts of the $\omega$ and $\rho$ mesons in the medium. The dominant 
background processes are seen to result from pion-nucleon ($\pi N$) 
bremsstrahlung, the $\eta$ Dalitz decay (denoted by $\eta$; solid 
line) and the $\omega$ Dalitz decay ($\omega \rightarrow \pi^0 
e^+e^-$). The $\Delta$ Dalitz decay ($\Delta$) as well as 
proton-neutron ($pn$) bremsstrahlung are of minor importance.  The 
upper solid curve represents the sum of all contributions.  We note 
that the width of the $\rho$--peak here is about 220 MeV, which is 
essentially larger than its vacuum width $\Gamma_{\rho}\approx$ 150 
MeV due to collisional broadening as described above. Nevertheless, 
above about 0.65 GeV of invariant mass $M$ the spectrum is fully 
dominated by the vector meson decays with a low background.

In Fig. 4 we show the resulting inclusive dilepton spectrum for the 
same system including collisional broadening as well as mass shifts 
of the $\omega$ and $\rho$ mesons in the medium according to Eq. 
(15). The background processes here are practically the same as in 
Fig. 3 and dominate for $M \leq$ 0.6 GeV.  When comparing the total 
spectrum from Fig. 3 with that from Fig. 4 we find an enhancement by 
a factor of 2 for 0.65 $\leq M \leq$ 0.75 GeV for the dropping vector 
meson masses as well a decrease of the spectrum for $M \geq$ 0.85 GeV 
due to the shifted $\rho$ mass because the $\rho$ almost completely 
decays inside the Pb-nucleus. The relative modifications of the 
dilepton spectra are qualitatively very similar to the situation at 
SPS energies in case of nucleus-nucleus collisions \cite{brat}, when 
employing a mass resolution of $\Delta M$ = 10 MeV as in the present 
case. Thus dileptons from pion-nucleus reactions at much lower energy 
also qualify for the experimental investigation of in-medium vector 
meson properties.

In summary, we have presented first calculations for dilepton 
production in pion-nucleus reactions and investigated the various 
contributing channels as well as in-medium modifications of the 
vector mesons due to collisional broadening or in-medium mass shifts 
\cite{1,2,3,4}. Our results for $\pi^- + $ Pb at 1.3 GeV/c indicate 
that the dominant background for invariant masses $M$ above 0.6 GeV 
arises from $\pi^- N$ bremsstrahlung which, however, is still small 
compared to the yield from the direct vector meson decays.  A mass 
shift of the $\rho$ and $\omega$ mesons should be seen experimentally 
by an enhanced yield in the mass regime 0.65 $\leq M \leq$ 0.75 GeV 
and a mass shift of the $\rho$ meson especially for $M \geq $ 0.85 
GeV because the $\rho$ almost completely decays inside a Pb-nucleus. 
The in-medium modifications of the $\omega$ mesons are most 
pronounced for small momentum cuts on the $e^+e^-$ pair in the 
laboratory.  This, however, will require a high mass resolution at 
least in the order of the free $\omega$ width.

\vspace{1cm}
The authors acknowledge many helpful discussions with E.L. 
Bratkovskaya and A. Sibirtsev throughout this study.

\newpage
\centerline{\bf {Figure captions}}

\vspace{5mm}
\noindent
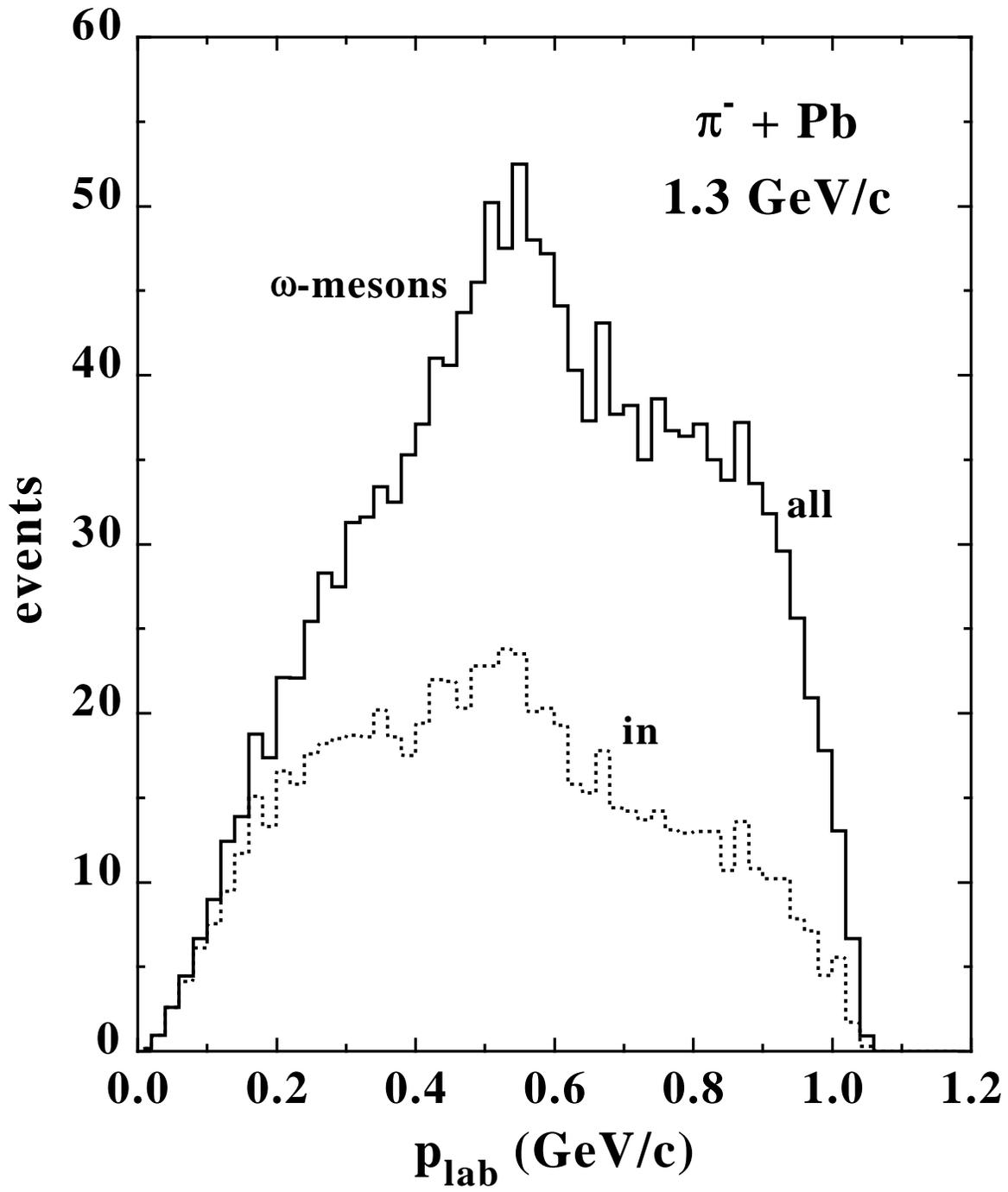
{\bf Fig.~1:} The momentum distribution of the $\omega$ mesons 
produced in the reaction $\pi^- Pb \to \omega X$\ at 1.3 GeV/c in the 
laboratory; solid histogram:   all $\omega$ mesons; dotted histogram: 
those decaying inside the nucleus.

\vspace{1cm}
\noindent
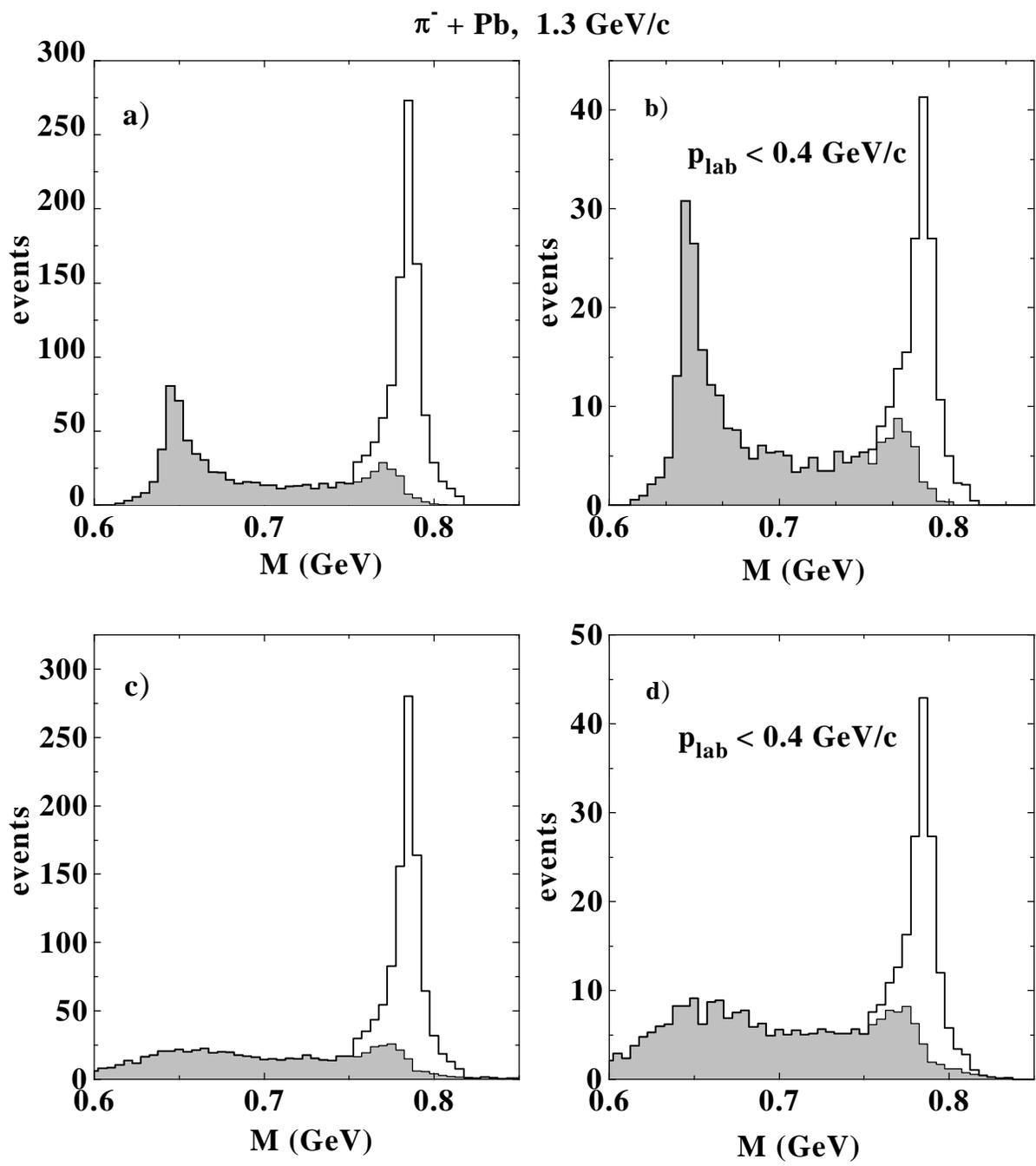
{\bf Fig.~2:} The $e^+e^-$ mass spectrum from $\omega$-decay when 
including a medium dependent $\omega$-mass shift according to Hatsuda 
and Lee (\cite{3}): $a)$ and $b)$ -  without collisional broadening; 
$c)$ and $d)$ -  with collisional broadening. Figs. $a)$ and $c)$ 
describe the full mass distributions, $b)$ and $d)$  events with a 
$\omega$ momentum cut $p_{lab} \leq 0.4 GeV/c$.  The hatched 
histograms describe the contribution of the '$in$' component, while 
the solid histograms display the total contribution from the '$in$' 
and '$out$' components, respectively.

\vspace{1cm}
\noindent
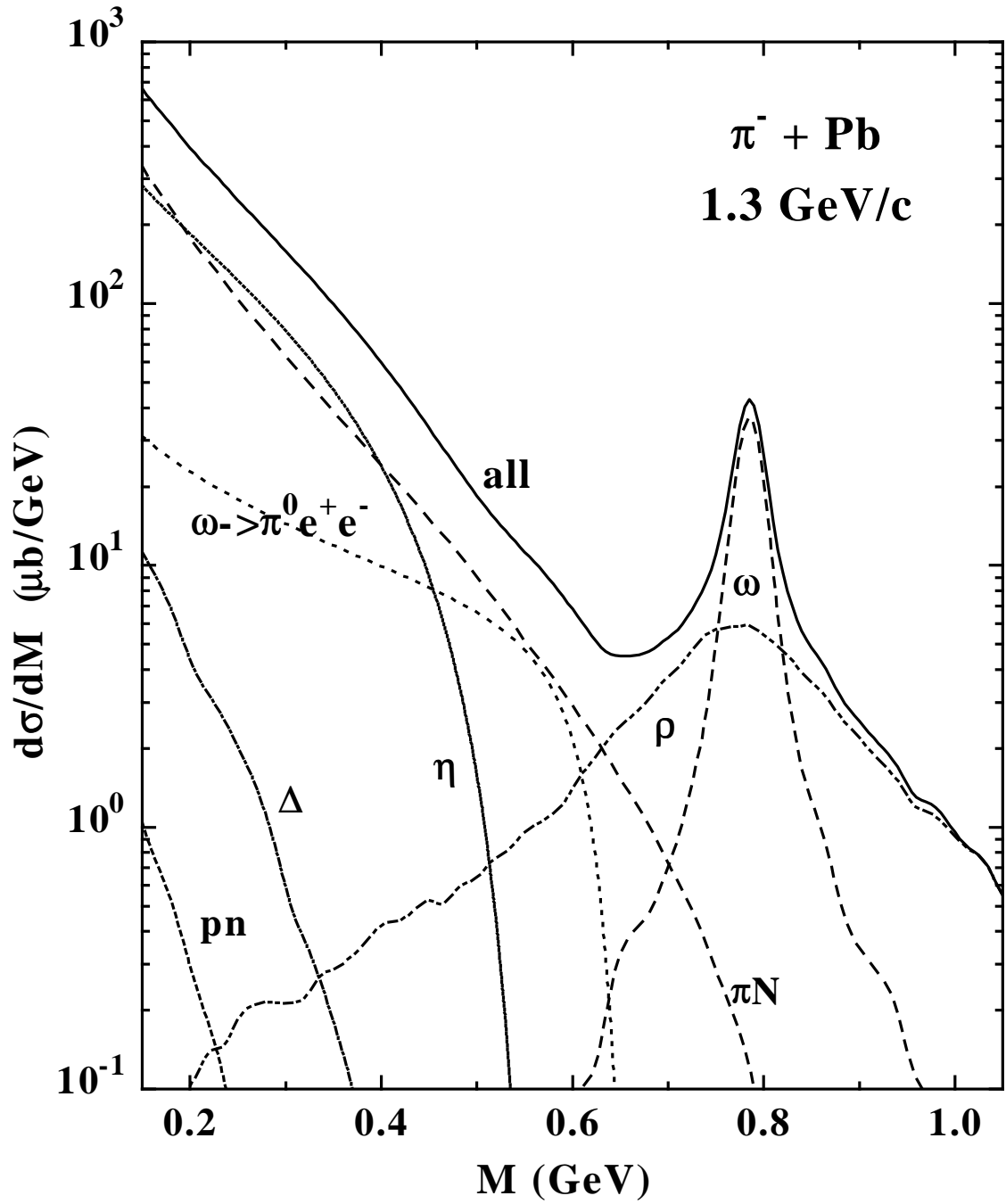
{\bf Fig.~3:} The dilepton spectra for $\pi^-$ + Pb reactions at 1.3 
GeV/c in logarithmic representation when  a mass shift of the vector 
mesons is neglected and collisional broadening  according to Eq. (12) 
is taken into account.  The upper solid curve is the sum of all 
contributions; the individual channels correspond to the $\eta$ 
Dalitz decay ($\eta$), pion-nucleon bremsstrahlung ($\pi N$), the 
$\omega$ Dalitz decay ($\omega \rightarrow \pi^0 e^+e^-$), the 
$\Delta$ Dalitz decay ($\Delta$), proton-neutron bremsstrahlung 
($pn$), and the direct decays of the vector mesons $\rho$ and 
$\omega$. The mass resolution adopted is $\Delta M$ = 10 MeV.

\vspace{1cm}
\noindent
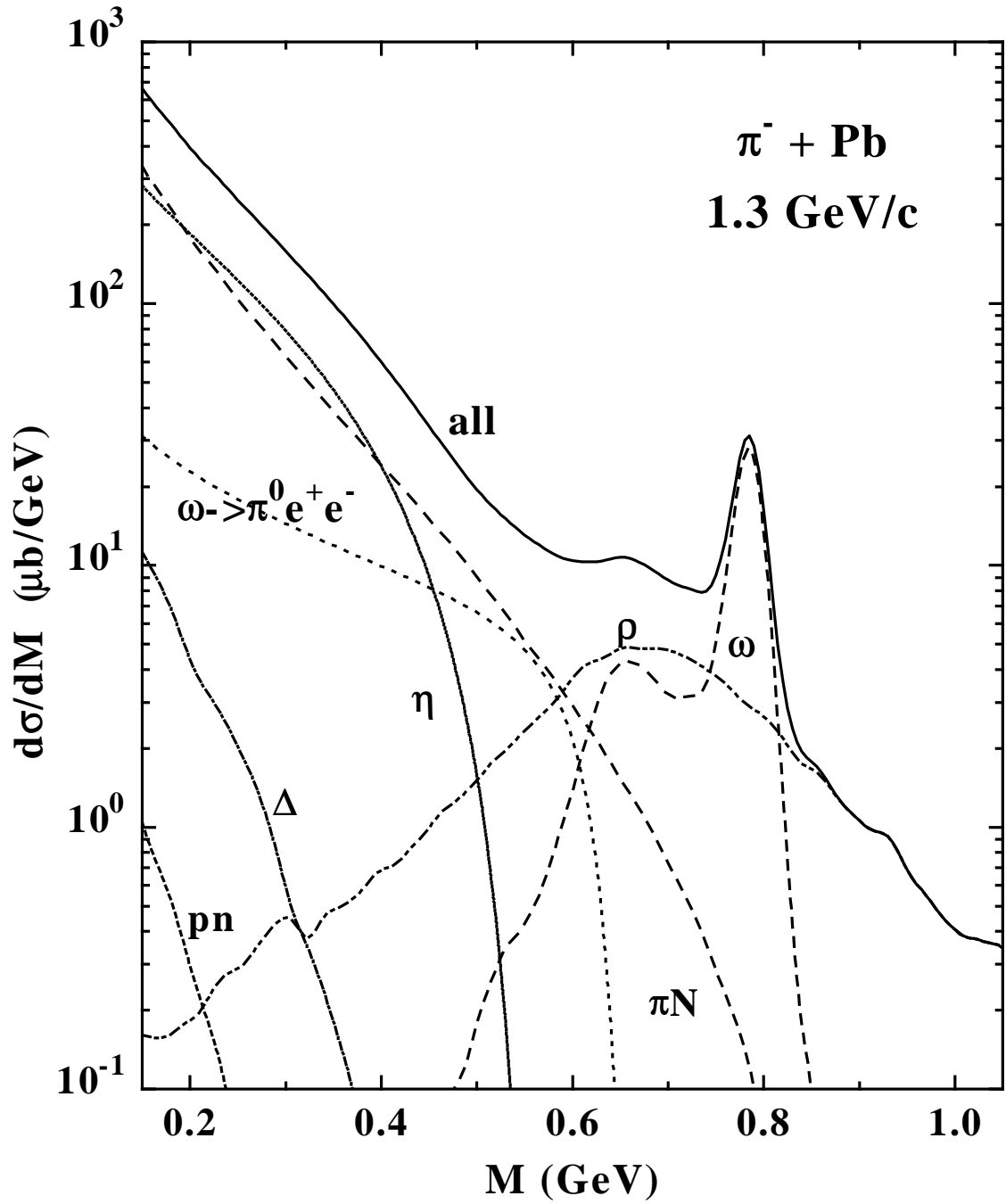
{\bf Fig.~4:} The dilepton spectra for $\pi^-$ + Pb reactions at 1.3 
GeV/c in logarithmic representation when including a mass shift of 
the vector mesons according to Eq. (15) and collisional broadening 
according to Eq. (12). The notations are the same as in Fig. 3.


\begin{thebibliography}{99}
\bibitem{1} G. Brown and M. Rho, Phys. Rev. Lett. 66 (1991) 2720.
\bibitem{2} C.M. Shakin and W.-D. Sun, Phys. Rev. C 49 (1994) 1185.
\bibitem{3} T. Hatsuda and S. Lee,  Phys. Rev. C 46 (1992) R34.
\bibitem{4} M. Asakawa and C.M. Ko, Phys. Rev. C 48 (1993) R526.
\bibitem{5} L. A. Kondratyuk, M. Krivoruchenko, N. Bianchi,
E. De Sanctis and V. Muccifora, Nucl. Phys. A 579 (1994) 453.
\bibitem{6} K.G. Boreskov, J. Koch, L.A. Kondratyuk and
 M.I. Krivoruchenko, Proc. of the 3rd International
Conference on Nucleon-Antinucleon Physics
(NAN'95), Phys. of Atomic Nuclei 10 (1996) 1908.
\bibitem{CERES} G. Agakichiev et al., Phys. Rev. Lett. 75 (1995) 1272.
\bibitem{HELIOS} M.A. Mazzoni, Nucl. Phys. A 566 (1994) 95c.
\bibitem{HELI} T. Akesson et al., Z. Phys. C 68 (1995) 47.
\bibitem{Li} G.Q. Li, C.M. Ko, and G.E. Brown, Phys. Rev. Lett. 75 (1995) 
4007.
\bibitem{11} W. Cassing, W. Ehehalt and C.M. Ko,
Phys. Lett B 363 (1995) 35.
\bibitem{12} W. Cassing, W. Ehehalt and I. Kralik,
Phys. Lett B 377 (1996) 5.
\bibitem{Wamb} R. Rapp, G. Chanfray, and J. Wambach,
Phys. Rev. Lett. 76 (1996) 368.
\bibitem{Metag} V. Metag, private communication.
\bibitem{7}  Ye.S. Golubeva, A.S. Iljinov, B.V. Krippa
and I.A. Pshenichnov, Nucl. Phys. A 537 (1992) 393.
\bibitem{8} Ye.S. Golubeva, A.S. Iljinov and L.A. Kondratyuk,
Proc. of the 3rd International
Conference on Nucleon-Antinucleon Physics
(NAN'95), Phys. of Atomic Nuclei No.10 (1996) 1894.
\bibitem{9} Ye.S. Golubeva, A.S. Iljinov and L.A. Kondratyuk,
Proc. of the 3rd International Conference "Mesons-96", Cracow, May 1996, 
to be published by World Scientific, Singapore, 1997.
\bibitem{10}  Gy. Wolf, G. Batko, W. Cassing et al., Nucl. Phys. A 517 (1990) 
615; \\ Gy. Wolf, W. Cassing and U. Mosel, Nucl. Phys. A 552 (1993) 549.
\bibitem{13} V. Eletsky and B.L. Ioffe, Preprint KFA-IKP(TH)-1996-10.
\bibitem{14} A. Sibirtsev, Nucl. Phys. A604 (1996) 455; \\
A. Sibirtsev, W. Cassing and U. Mosel, nucl-th/9607047, Z. Phys. A, in print
\bibitem{15} Ye.S. Golubeva, A.S. Iljinov and I.A. Pshenichnov,
Nucl. Phys. A 562 (1993) 389
\bibitem{brat} E. L. Bratkovskaya and W. Cassing, nucl-th/9611042, submitted
 to Nucl. Phys. A.
\end{thebibliography}
\end{document}